\documentstyle{article}
\def\bp#1{\overline{#1}^\prime}

\begin{document}

\hfill{UM-P-97/21}

\vspace{30mm}
\begin{center}
{\LARGE \bf Fermion Masses and Mixing in 331 Models with
Horizontal Symmetry}

\vspace{30mm}
{\bf M.B.Tully and G.C.Joshi}

\vspace{10mm}
{\sl 
Research Centre for High Energy Physics \\
School of Physics, University of Melbourne, \\
Parkville, Victoria 3052, Australia.}

\end{center}

\vspace{40mm}

\begin{abstract}

The possibility of adding an $SU(2)$ horizontal symmetry to
the 331 model is studied. It is found that simple, anomaly-free 
fermion assignments can be made which lead to plausible 
results for fermion masses and mixings. In particular, all
particles of the first generation are massless at tree-level,
and the CKM matrix acquires a realistic form. 
\end{abstract}

\newpage

\section{Introduction}

Two of the most striking inadequacies of the Standard Model 
are its inability to account for the fermion masses and for the
existence of a family structure, whereby there exist three families
of particles with the same quantum numbers but different masses. 
These facts are accomodated but not in any way explained.
In the 331 model, however, introduced by Pisano, Pleitez and 
Frampton  \cite{PIS1} -- \cite{OZE1} , and so-called because the gauge group is
$SU(3)_C \times SU(3)_L \times U(1)_X$, the number of families
is required to be a multiple of three to cancel anomalies. It
is a feature of this type of model that the third family is 
treated differently to the first two. \cite{LIU1} Anomalies do not
cancel within each family as is the case with the Standard
Model and most extensions to it, but rather, only cancel when the
contributions from all three families are taken into account.
In this sense, the model can be said to be addressing the 
reason for the existence of three families.

However, it is still the case in this type of model 
that the fermion masses depend on arbitrary Yukawa coupling
strengths. In the past, many attempts have been made 
to reduce some of this arbitrariness in the Standard Model 
by postulating the
existence of a horizontal symmetry, by which is meant a
symmetry involving the corresponding particles of different
families rather than the different particles of the same
family. One of the popular choices for this symmetry has been
$SU(2)$, and this has been extensively studied in
terms of what effect the existence of such a symmetry would
have on the relations between the fermion masses and also
the Cabibbo-Kobayashi-Maskawa (CKM) matrix. \cite{SHA1} 

In references \cite{PLE1} and \cite{PLE2}
Pleitez made the suggestion of adding such an $SU(2)$ horizontal
symmetry to the 331 model, and this is what has been pursued
in this paper, investigating the effects on fermion masses and
mixings. We consider both the original 331 model (Model I) and
the later extension containing right-handed neutrinos (Model II).
In both cases, fermions from the first two families will transform
as doublets under the horizontal symmetry, and those from the third
as singlets, following the general pattern of the 331-model.
\footnote{Such an arrangment in the context of the Standard Model
has also been tried and gives similar results to the ones obtained 
here \cite{FOO2}}
It will be shown below that the resulting mass spectrum is a very plausible
one, and contains the following features: 

- the electron, the electron neutrino (in some cases), the up quark 
and the down quark are massless at tree-level, while all other
fermions do have mass.

- the CKM matrix has the form (again, at tree-level):
\begin{equation}
U_{CKM}=\left( \begin{array}{ccc} \cos \theta & \sin \theta & 0 \\
-\sin \theta & \cos \theta & 0 \\ 0 & 0 & 1 \end{array} \right)
\label{eq:CKM}
\end{equation}

Clearly, such results closely resemble the form of the 
actual data. It is proposed that radiative corrections
would then lead to small values for the masses of the first
family, and the quantitites in the CKM matrix shown 
above as zero.

\section{Model I}
In this section, we will consider exactly the same fermion
content as contained in the original 331 model, but with the
particles also transforming under the $SU(2)$ horizontal symmetry
in the manner described in the introduction, namely, the first
two familes in doublets and the third in singlets. This arrangement
satisfies the requirement that the gauge anomaly, 
$\left[ SU(2)_H \right]^2 U(1)_Y$ vanishes, and indeed, is the
only possible assignment where this is true and all fermions of
each family transform the same way. 
(It should be noted, however, that this assignment does not cancel the
global $SU(2)$ anomaly, as the number of doublets is odd. A
possible remedy for this is to include right-handed neutrinos
transforming as singlets under $SU(3)_L$, and this will be considered
at the end of this section).

The fermion assignments under $SU(3)_C \times SU(3)_L \times SU(2)_H
\times U(1)_X$ are as follows:
\begin{eqnarray}
f_L=\left( \begin{array}{ll} e^- & \mu^- \\ \nu_e & \nu_{\mu} \\ 
e^+ & \mu^+ \end{array} \right)_L & \sim
 (1,3^\star,2,0) & \\
f_{3L}=\left( \begin{array}c \tau^- \\ \nu_{\tau} \\ \tau^+ \end{array} 
\right)_L & \sim (1,3^\star,1,0) & \\
q_L= \left( \begin{array}{cc} u & c \\ d & s \\ D_1 & D_2 \end{array} 
\right)_L & \sim (3,3,2,-\frac{1}{3}) \\
q_{3L} = \left( \begin{array}c b \\ t \\ T \end{array} \right)_L 
& \sim (3,3^\star,1,+\frac{2}{3}) \\
u_R= \left( u,c \right)_R & \sim (3,1,2,+\frac{2}{3}) \\
u_{3R}=t_R & \sim (3,1,1,+\frac{2}{3}) \\
d_R= \left( d,s \right)_R & \sim (3,1,2,-\frac{1}{3}) \\
d_{3R}= b_R & \sim (3,1,1,-\frac{1}{3}) \\
D_R= \left( D_1,D_2 \right)_R & \sim (3,1,2,-\frac{4}{3}) \\
 T_R & \sim (3,1,1,+\frac{5}{3})
\end{eqnarray}
Note that $D$ and $T$ denote exotic quarks with charge $-\frac{4}{3}$ and
$+\frac{5}{3}$ respectively.

As in the minimal-331 model, a total of three Higgs triplets and one sextet
is required, all of which transform as doublets under the horizontal
symmetry except for one, $\chi$, which transforms as a singlet.
\begin{eqnarray}
\phi= \left( \begin{array}l \phi^0_i \\ \phi^-_i \\ \phi^{--}_i \end{array} \right) & 
\sim (1,3,2,-1) & i=1,2\\
\eta= \left( \begin{array}l \eta^+_i \\ \eta^0_i \\ \eta^-_i \end{array}
\right) & \sim (1,3,2,0) & \\
S= \left( \begin{array}{lll} s^{--}_i & s^{-}_i & s^{0}_i \\ s^{-}_i & s^{0}_i &
s^{+}_i \\ s^{0}_i & s^{+}_i & s^{++}_i \end{array} \right) & \sim (1,6,2,0)
\\
\chi= \left( \begin{array}l \chi^{++} \\ \chi^{+} \\ \chi^{0} \\ 
\end{array} \right) & \sim (1,3,1,+1)
\end{eqnarray}

The Yukawa interaction (in an abbreviated form omitting $SU(3)$ and
$SU(2)$ indices) is then:
\begin{equation}
{\cal{L}}^{Yukawa}_{leptons}  =  \lambda_1 \overline{f}_L \left( f_{3L} \right)^c 
S + \lambda_2 \overline{f}_L \left( f_{3L} \right)^c \eta
\end{equation}
\begin{equation} 
{\cal{L}}^{Yukawa}_{quarks}  =  \lambda_3 \overline{q}_L u_{3R} \phi + 
\lambda_4 \overline{q}_L d_{3R} \eta +
\lambda_5 \overline{q}_{3L} u_R \eta^\star + 
\lambda_6 \overline{q}_{3L} d_R \phi^{\star} + 
\lambda_7 \overline{q}_L D_{R} \chi + 
\lambda_8 \overline{q}_{3L} T_{R} \chi^{\star} \label{eq:coup}
\end{equation}

To study the effects of symmetry breaking, we will now
study this Lagrangian in more detail.
Writing out two of the terms in equation \ref{eq:coup} in full as
an example, we have:
\begin{equation}
\lambda_3 Tr \left[ \left( \begin{array}{ccc} \bp{u} & \bp{d} 
& \bp{D}_1 \\ 
\bp{c} & \bp{s} & \bp{D_2} \end{array} \right) \left( \begin{array}{ll} 
\phi_1^0 & \phi_2^0 \\ 
\phi_1^- & \phi_2^- \\ \phi_1^{--} & \phi_2^{--} \end{array} \right) \right] t_R^\prime +
\lambda_5 \left( \bp{b},\bp{t},\bp{T} \right) 
\left( \begin{array}{ll} \eta_1^- & \eta_2^- \\ \overline{\eta}_1^0 & 
\overline{\eta}_2^0 \\ 
\eta_1^+ & \eta_2^+ \end{array} \right) 
\left( \begin{array}{c} u^\prime \\ c^\prime  \end{array} \right)_R
\end{equation}
(where the primes denote symmetry eigenstates) and the other terms follow similarly. 
It is then easily seen that when $\phi$ and $\eta$ develop vacuum
expectation values 

\begin{eqnarray}
v_i= \langle \phi_i^0 \rangle  & i=1,2 \\
w_i= \langle \eta_i^0 \rangle  & \\
\end{eqnarray}

the following mass terms arise.

\begin{equation}
{\cal{L}}^{quark}_{mass}= \overline{U}_L^\prime M^U U_R^\prime 
+ \overline{D}_L^\prime M^D D_R^\prime
\end{equation}
where
\begin{equation}
U^\prime=\left( \begin{array}{c} u^\prime \\ c^\prime \\ t^\prime  \end{array} \right) , 
D^\prime=\left(
\begin{array}{c} d^\prime \\ s^\prime \\ b^\prime \end{array} \right)
\end{equation}
and the mass matrices for the charge $\frac{2}{3}$ and charge $\frac{1}{3}$
quarks are given by:
\begin{equation}
M^U  =  \left( \begin{array}{ccc} 0 & 0 & \lambda_3 v_1 \\ 0 & 0 &
\lambda_3 v_2 \\ \lambda_5 w_1 & \lambda_5 w_2 & 0 \end{array} \right)
\end{equation}
and
\begin{equation}
M^D  =  \left( \begin{array}{ccc} 0 & 0 & \lambda_4 w_1 \\ 0 & 0 &
\lambda_4 w_2 \\ \lambda_6 v_1 & \lambda_6 v_2 & 0 \end{array} \right)
\end{equation}

In order to find the mass eigenstates, these matrices
are diagonalised by introducing the unitary matrices $A_{L,R}$
and $B_{L,R}$, such that:

\begin{equation}
U^\prime_L=A_L U_L, U^\prime_R=A_R U_R, D^\prime_L=B_L D_L, D^\prime_R=
B_R D_R
\end{equation}
and
\begin{eqnarray}
A_L^\dagger M^U A_R = D^U \\
B_L^\dagger M^D B_R = D^D
\end{eqnarray}
For the charge $\frac{2}{3}$ quarks,
\begin{equation}
A_L=\frac{1}{\sqrt{v_1^2+v_2^2}} \left( \begin{array}{rcc} -v_2 & v_1 &
0 \\ v_1 & v_2 & 0 \\ 0 & 0 & \sqrt{v_1^2 + v_2^2} \end{array} \right)
\end{equation}
\begin{equation}
A_R=\frac{1}{\sqrt{w_1^2+w_2^2}} \left( \begin{array}{rcc} -w_2 & 0 & w1
\\ w_1 & 0 & w_2 \\ 0 & \sqrt{w_1^2+w_2^2} & 0 \end{array} \right)
\end{equation}
to give
\begin{equation}
D^U  =  \left( \begin{array}{ccc}  0 & 0 & 0 \\ 0 & \lambda_3
(v_1^2+v_2^2)^{\frac{1}{2}} & 0 \\ 0 & 0 & \lambda_5 (w_1^2 +
w_2^2)^{\frac{1}{2}} \end{array} \right)
\end{equation}
while for the charge $\frac{1}{3}$ quarks,
\begin{equation}
B_L= \frac{1}{\sqrt{w_1^2+w_2^2}} \left( \begin{array}{rcc} -w_2 & w_1 &
0 \\ w_1 & w_2 & 0 \\ 0 & 0 & \sqrt{w_1^2+w_2^2} \end{array} \right)
\end{equation}
\begin{equation}
B_R= \frac{1}{\sqrt{v_1^2+v_2^2}} \left( \begin{array}{rcc} -v_2 & 0 
& v_1 \\ v_1 & 0 & v_2 \\ 0 & \sqrt{v_1^2+v_2^2} & 0 \end{array} \right)
\end{equation}
giving:
\begin{equation}
D^D  =  \left( \begin{array}{ccc} 0 & 0 & 0 \\ 0 & \lambda_4(w_1^2 +
w_2^2)^{\frac{1}{2}} \\ 0 & 0 & \lambda_6 (v_1^2+v_2^2)^{\frac{1}{2}}
\end{array} \right) 
\end{equation}
The CKM matrix (denoted $U_{CKM}$) is given
by $A_L^\dagger B_L$ and is therefore:
\begin{equation}
U_{CKM}=\frac{1}{\sqrt{(v_1^2+v_2^2)(w_1^2+w_2^2)}} \left(
\begin{array}{rrr} v_1w_1+v_2w_2 & v_1 w_2 - v_2 w_1 & 0 \\
-v_1 w_2 + v_2 w_1 & v_1 w_1 + v_2 w_2 & 0 \\
0 & 0 & \sqrt{(v_1^2+v_2^2)(w_1^2+w_2^2)} \end{array} \right)
\end{equation}
which can clearly be re-parameterised in the form of equation \ref{eq:CKM}
by setting 
\begin{equation}
\tan \theta = \frac{v_1w_2-v_2w_1}{v_1w_1+v_2w_2}
\end{equation}
The exotic quarks $D_1$,$D_2$ and $T$ acquire masses $\lambda_7 u$,
$\lambda_7 u$ and $\lambda_8 u$ respectively, where
\begin{equation}
u=\langle \chi^0 \rangle
\end{equation}

In the lepton sector, as in the usual 331 model both a triplet
and a sextet are required to produce a realistic mass spectrum.
This is because a triplet by itself would lead to an anti-symmetric
mass matrix in flavour space (with eigenvalues $0$,$+M$ and $-M$)
\cite{FOO1}, while a sextet leads to a symmetric mass matrix.

The coupling of the leptons to the Higgs triplet is given by:
\begin{equation}
\epsilon_{ijk} \overline{f}^i_{La} \left( f^j_{3L} \right)^c \eta^{ka}
\end{equation}
where $i$,$j$ and $k$ are $SU(3)$ indices and $a$ an $SU(2)$ index.
Writing, as earlier, $w_i=\langle \eta_i^0 \rangle$ the following
mass matrix for the charged leptons develops:
\begin{equation}
M^E_{triplet}=\lambda_2 \left( \begin{array}{ccc} 0 & 0 & -w_1 \\
0 & 0 & -w_2 \\ w_1 & w_2 & 0 \end{array} \right)
\end{equation}
The sextet coupling is given by:
\begin{equation}
\overline{f_a^i} \left( f_3^j \right) ^c S^a_{ij}
\end{equation}
so when $S$ develops a VEV
\begin{equation}
s_i= \langle S^{13}_i \rangle
\end{equation}
a symmetric mass matrix arises:
\begin{equation}
M^E_{sextet}= \lambda_1 \left( \begin{array}{ccc} 0 & 0 & s_1 \\ 
0 & 0 & s_2 \\ s_1 & s_2 & 0 \end{array} \right)
\end{equation}
Adding these gives the total mass matrix for the charged leptons.
\begin{equation}
M^E= \left( \begin{array}{ccc} 0 & 0 & \lambda_1s_1 - \lambda_2 w_1 \\
0 & 0 & \lambda_1 s_2 - \lambda_2 w_2 \\ \lambda_1 s_1 + \lambda_2 w_1 &
\lambda_1 s_2 + \lambda_2 w_2 & 0 \end{array} \right)
\end{equation}
which can be diagonalised to give the mass eigenvalues.
\begin{equation}
D^E  =  \left( \begin{array}{ccc} 0 & 0 & 0 \\ 0 & \left[(\lambda_1 s_1 -
\lambda_2 w_1)^2 + (\lambda_1 s_2 - \lambda_2 w_2)^2 \right]^{\frac{1}{2}}
 & 0 \\ 0 & 0 & \left[ (\lambda_1 s_1 + \lambda_2 w_1)^2 + (\lambda_1 s_2 
+ \lambda_2 w_2)^2 \right] ^{\frac{1}{2}}
\end{array} \right) 
\end{equation}

Notice that, as in the quark sector, the electron is
massless at the tree-level, while the two heavier leptons 
do develop masses. 

To summarise, then, the nine fermion masses are given by:
\begin{equation}
\begin{array}{ccc}
m_u=0 & m_d=0 & m_e=0 \\
m_c=\lambda_3 \left( v_1^2 + v_2^2 \right) ^{\frac{1}{2}}  &
m_s=\lambda_4 \left( w_1^2 + w_2^2 \right) ^{\frac{1}{2}} &
m_{\mu}= \left[ (\lambda_1 s_1 - \lambda_2 w_1)^2 + (\lambda_1 s_1
- \lambda_2 w_2)^2 \right] ^\frac{1}{2} \\
m_t=\lambda_5 (w_1^2 + w_2^2 )^{\frac{1}{2}} &
m_b=\lambda_6 (v_1^2 + v_2^2 )^{\frac{1}{2}} &
m_{\tau}= \left[ (\lambda_1 s_1 + \lambda_2 w_1)^2 + (\lambda_1 s_1
+\lambda_2 w_2)^2 \right] ^\frac{1}{2} 
\end{array} \label{eq:SPEC1}
\end{equation}

Right-handed neutrinos could also be added to this model in a
straightforward manner. Under $SU(3)_C \times SU(3)_L \times SU(2)_H
\times U(1)_X$ they transform as:
\begin{eqnarray}
\nu_R= \left( \nu_e,\nu_\mu \right)_R & \sim (1,1,2,0) \\
\nu_{3R}= \nu_{\tau R} & \sim (1,1,1,0)
\end{eqnarray}
and couple to the Higgs triplet $\eta$ according to:
\begin{equation}
{\cal{L}}^{Yukawa}_{neutrinos}=\lambda_9 \overline{f}_L \nu_{3R}\eta^\star
+ \lambda_{10} \overline{f}_{3L}\nu_R \eta^\star
\end{equation}
Their mass matrix is therefore:
\begin{equation}
M^\nu= \left( \begin{array}{ccc} 0 & 0 & -\lambda_9 w_2 \\ 0 & 0 &   
\lambda_9 w_1 \\ \lambda_{10} w_1 & \lambda_{10} w_2 & 0 \end{array}
\right)
\end{equation}
which is diagonalised to:
\begin{equation}
D^\nu= \left( \begin{array}{ccc} 0 & 0 & 0 \\ 0 & \lambda_9 
\sqrt{w_1^2+w_2^2} & 0\\
0 & 0 & \lambda_{10} \sqrt{w_1^2+w_2^2} \end{array} \right)
\end{equation}

\section{Model II}
In this section, an SU(2) horizontal symmetry will be added to
the 331 model extended to include right-handed
neutrinos in the lepton triplet. \cite{MON1,FOO3,LON1,LON2} 
The right-handed electron
now transforms as an $SU(3)_L$ singlet. This model has the advantage 
compared to the original
331 model of only needing three Higgs triplets to give masses to
the fermions, rather than the three triplets and a sextet.
The three exotic quarks are still present but now have charge
$- \frac{1}{3}$ and $\frac{2}{3}$ and will be denoted $D_1$,
$D_2$ and $U$.

As in Model I, the first two generations of fermions will transform
as doublets under the horizontal symmetry and the third generation
as singlets. This arrangement can be shown to cancel all $\left[SU(2)_H 
\right]^2 U(1)_X$ anomalies. Although there are a number of other
possible combinations of fermion assignments, again as in Model I they all
suffer from having fermions of the same family being placed in 
different $SU(2)_H$ representations. This arrangement also cancels
the global $SU(2)$ anomaly.

Thus, the fermion assignments under $SU(3)_C \times SU(3)_L \times
SU(2)_H \times U(1)_X$ are as follows:

\begin{eqnarray}
f_L= \left( \begin{array}{ll} \nu_e & \nu_{\mu} \\ e^- & \mu^- \\
\nu^c_e & \nu^c_{\mu} \end{array} \right) _L & \sim (1,3,2,-\frac{1}{3}) \\
f_{3L} = \left( \begin{array}{l} \nu_{\tau} \\ \tau^- \\ \nu^c_{\tau} 
\end{array} \right) _L & \sim (1,3,1,-\frac{1}{3}) \\
e_R=\left( e^-,\mu^- \right)_R & \sim (1,1,2,-1) \\
e_{3R}= \tau^-_R & \sim (1,1,1,-1) \\
q_L = \left( \begin{array}{ll} d & s \\ u & c \\ D_1 & D_2 \end{array}
\right)_L & \sim (3,3^{\star},2,0) \\
q_{3L} = \left( \begin{array}{l} t \\ b \\ U \end{array} \right)
& \sim (3,3,1,-\frac{1}{3}) \\
u_R = \left( u,c \right)_R & \sim (3,1,2,+\frac{2}{3}) \\
u_{3R}=t_R & \sim (3,1,1,+\frac{2}{3}) \\
d_R = \left( d,s \right)_R & \sim (3,1,2,-\frac{1}{3}) \\
d_{3R} = b_r & \sim (3,1,1,-\frac{1}{3}) \\
D_R = \left( D_1 , D_2 \right)_R & \sim (3,1,2,-\frac{1}{3}) \\
T_R & \sim (3,1,1,+\frac{2}{3})
\end{eqnarray}

The three Higgs triplets transform as two doublets and a singlet
under $SU(2)_H$.

\begin{eqnarray}
\rho= \left( \begin{array}l \rho^+_i \\ \rho^0_i \\ \rho^{\prime +}_i \end{array}
\right) & \sim (1,3,2,+\frac{2}{3}) & i=1,2 \\
\eta= \left( \begin{array}l \eta^0_i \\ \eta^-_i \\ \eta^{\prime 0}_i \end{array}
\right) & \sim (1,3,2,-\frac{1}{3}) & \\
\chi= \left( \begin{array}l \chi^+ \\ \chi^0 \\ \chi^{\prime +} \end{array}
\right) & \sim (1,3,1,+\frac{2}{3}) & 
\end{eqnarray}

The Yukawa couplings for the leptons and quarks are then given by
\begin{equation}
{\cal{L}}^{Yukawa}_{leptons}=
\lambda_1 \overline{f}_Le_{3R} \rho + \lambda_2 \overline{f}_{3L} e_R \rho
+ \lambda_3 \overline{f}_L \left( f_{3L} \right) ^c \rho^{\star}
\end{equation}
\begin{equation}
{\cal{L}}^{Yukawa}_{quarks}=
\lambda_4 \overline{q}_L u_{3R} \rho^{\star} + 
\lambda_5 \overline {q}_L d_{3R} \eta^{\star} + 
\lambda_6 \overline{q}_{3L} u_R \eta + 
\lambda_7 \overline{q}_{3L} d_R \rho + 
\lambda_8 \overline{q}_L D_R \chi^{\star}+ 
\lambda_9 \overline{q}_{3L} T_R \chi
\end{equation}
When the Higgs fields develop VEVs:
\begin{eqnarray}
v_i=\langle \rho^0_i \rangle & i=1,2 \\
w_i=\langle \eta^0_i \rangle & \\
u_i=\langle \chi^0 \rangle & 
\end{eqnarray}
the quarks obtain masses of the same form as in Model I, which are 
given in equation \ref{eq:SPEC2} below. The lepton masses, however
now develop somewhat differently.

The charged leptons acquire a mass matrix
\begin{equation}
M^E= \left( \begin{array}{ccc} 0 & 0 & \lambda_1 v_1 \\ 0 & 0 & 
\lambda_1 v_2 \\ -\lambda_2 v_2 & \lambda_2 v_1 & 0 \end{array} \right)
\end{equation}
which can be diagonalised to give the mass eigenstates as follows:
\begin{equation}
D^E= \left( \begin{array}{ccc} 0 & 0 & 0 \\ 0 & \lambda_1 \sqrt{v_1^2 + v_2^2} & 0 \\
0 & 0 & \lambda_2 \sqrt{v_1^2+v_2^2} \end{array} \right) 
\end{equation}
The neutrinos, on the other hand, develop an anti-symmetric mass
matrix:
\begin{equation}
M^{\nu}= \lambda_3 \left( \begin{array}{ccc} 0 & 0 & v_1 \\
0 & 0 & v_2 \\ -v_1 & -v_2 & 0 \end{array} \right)
\end{equation}
which when diagonalised, leads to one of the neutrinos being massless
at tree-level, with the other two having degenerate masses.
\begin{equation}
D^{\nu}= \left( \begin{array}{ccc} 0 & 0 & 0 \\ 0 & \lambda_3 \sqrt{v_1^2+v_2^2} & 0 \\
0 & 0 & \lambda_3 \sqrt{v_1^2+v_2^2} \end{array} \right)
\end{equation}

For Model II then, the summary of fermion masses is as follows:
\begin{equation}
\begin{array}{cccc}
m_u=0 & m_d=0 & m_e=0 & m_{\nu_e}=0 \\
m_c=\lambda_4 \left( v_1^2+v_2^2 \right) ^\frac{1}{2} & 
m_s = \lambda_5 \left( w_1^2+w_2^2 \right)^\frac{1}{2} & 
m_{\mu}=\lambda_1 \left(v_1^2+v_2^2 \right)^ \frac{1}{2} & 
m_{\nu_{\mu}}=\lambda_3 \left( v_1^2+v_2^2 \right)^\frac{1}{2} \\
m_t= \lambda_6 \left( w_1^2+w_2^2 \right)^\frac{1}{2} & 
m_b= \lambda_7 \left( v_1^2+v_2^2 \right)^\frac{1}{2} & 
m_\tau=\lambda_2 \left(v_1^2+v_2^2 \right)^\frac{1}{2} &
m_{\nu_\tau}= \lambda_3 \left( v_1^2+v_2^2 \right)^\frac{1}{2}
\end{array} \label{eq:SPEC2}
\end{equation}

\section{Conclusion}
We have shown that it is possible to extend the 331 model by adding an $SU(2)$ 
horizontal symmetry, and that simple choices of fermion assignments can be made
which not only cancel anomalies, but also lead to very plausible patterns
for fermion masses and mixings, as given in equations \ref{eq:CKM}, \ref{eq:SPEC1}
and \ref{eq:SPEC2} .                                           

It remains now to investigate other issues arising from this model such as the
existence of Flavour-Changing Neutral Currents and the details of the radiative
corrections needed to give mass to the fermions of the first family.

\end{document}